# Unique and Minimum Distance Decoding of Linear Codes with Reduced Complexity

D. Spasov (spasov.dejan@gmail.com)

*Abstract*—We show that for (systematic) linear codes the time complexity of unique decoding is $O\left(n^2 q^{nRH(\delta/2/R)}\right)$ and the time complexity of minimum distance decoding is $O\left(n^2 q^{nRH(\delta/R)}\right)$. The proposed algorithm inspects all error patterns in the information set of the received message of weight less than $d/2$ or $d$, respectively.

## I. Introduction

LET $C$ is a systematic linear code with parameters $[n,k,d]$. It is well known that Hamming balls $Ball(c,t)$ with radius $t = \lfloor(d-1)/2\rfloor$ around the codewords $c \in C$ are disjoint. Let $y$ is the received message. Then the Unique Decoding strategy is to find the codeword $c_y \in C$, such that $y \in Ball(c_y,t)$, or return incomplete decoding, i.e. $y \notin Ball(c,t) \forall c \in C$. Trivial way to do this is to inspect all $q^k$ codewords and return $c_y$ such that $d(y,c_y) \leq t$. The time complexity of this approach is $O(nq^{Rn})$. Another alternative for unique decoding, with time complexity $O(nq^{H(\delta/2)n})$, is to inspect all $V(n,t)$ error patterns $e$ and find the pattern such that $y-e$ is a codeword. Minimum Distance Decoding, on the other hand, inspects all $q^k$ codewords and returns $c_y \in C$ such that $d(y,c_y)$ is minimal or it inspects all error patterns of weight less than the covering radius [1].

In the following section we will show that unique decoding can be done by inspecting all $V(k,t)$ error patterns in the information set of the received message $y$. Then we will generalize this algorithm to perform Minimum Distance Decoding.

## II. The Algorithm

We will use $\langle a|b \rangle$ to denote concatenation of two vectors, such that $a$ belongs to the information set and $b$ belongs to the check set of a codeword. Let the message $x$ is encoded in the codeword $c_x = \langle x|r \rangle$ and sent over a noisy channel. Let the random error pattern is denoted with $e = \langle v|u \rangle$ and the received word is denoted with $y = \langle y_x|y_r \rangle$. The unique decoding algorithm, below, inspects all error patterns in the information set $e = \langle v|0 \rangle$ with weight $wt(e) \leq t$ and outputs the message $\langle x|u \rangle$ if $y$ belongs to some $Ball(c,t)$:

```
Unique_Decoding(y)
    Let  t=⌊(d-1)/2⌋
    1. find the syndrome s_y=Hy^T
    2. if wt(s_y)≤t return y
    3. foreach vector e=<v|0> s.t. wt(e)≤t
        a. find s_e=He^T
        b. if wt(e)+wt(s_y-s_e)≤t
                return y-e
    4. return -1 // incomplete decoding
```

*Proposition 1:* The `Unique_Decoding(y)` algorithm can remove any error pattern of weight $\leq t$ from the received message $y$.

*Proof:* Let $e_v = \langle v|0 \rangle$, $wt(e_v) \leq t$, is the coset leader and $s_v = He_v^T$ is the syndrome of a coset. Let assume that the pairs $(s_v, e_v)$ are explicitly known; for example, they are stored in a look-up table.

We will consider the error pattern $e = \langle v|u \rangle$ as a linear combination of two vectors

$$e = \langle v|0 \rangle + \langle 0|u \rangle = e_v + e_u \qquad (1)$$

Since $wt(e_u) \leq t$ and $s_u = He_u^T = u$, we can say that $e_u$ is

the leader and $u$ is the syndrome of the same coset. Hence, the syndrome $s$ of the received message $y$ is

$$s = Hy^T = H(c_x + e_v + e_u)^T = s_v + u \qquad (2)$$

From (2), we can formulate the decoding strategy: for each $e_v$ in the table $(s_v, e_v)$ denote with $x = y - e_v$ and compute the syndrome

$$s_x = Hx^T \qquad (3)$$

If

$$wt(e_v) + wt(s_x) \leq t \qquad (4)$$

then the error pattern that corrupted the message is

$$e = \langle e_v | s_x \rangle$$

∎

In worst case, the algorithm will check all $V_q(k,t)$ error patterns. So the time complexity is upper-bounded by

$$O\left(n^2 q^{RH\left(\frac{\delta}{2R}\right)n}\right)$$

If we use the fact that for long random linear codes the covering radius is equal to $d$, where $d$ is the largest integer solution of the Gilber-Varshamov inequality [2]. Then we can formulate Minimum Distance Decoding algorithm that inspects all error patterns of weight less than $d$ in the information set:

```
MD_Decoding(y)
    error_vector=0, error_wt=n
    1. compute the syndrome s_y=Hy^T
    3. foreach vector e=<v|0>, wt(e)≤d
      a. Find s_e=He^T
      b. if wt(e)+wt(s_y-s_e) ≤ error_wt
         i. error_wt=wt(e)+wt(s_y-s_e)
         ii. error_vector=e
    4. return y-error_vector
```

The proof of correctness of the above algorithm is similar to the proof of proposition 1. Thus the time complexity of MDD decoding is

$$O\left(n^2 q^{RH\left(\frac{\delta}{R}\right)n}\right)$$

and this result improves some of the previously known bounds on MD decoding found in [1], [2], [3], or [4].

On the other hand, pairs $(s_v, e_v)$ need not to be stored in a look-up table, but they can be listed by divide-and-conquer strategy in the course of decoding. Therefore, the space complexity of `Unique_Decoding(y)` and `MD_Decoding(y)` is proportional with the dimension of the generator matrix, i.e. $O(n^2)$.